\documentstyle[12pt,fleqn,epsf,epsfig]{article}
\def\lsim{\raise0.3ex\hbox{$\;<$\kern-0.75em\raise-1.1ex\hbox{$\sim\;$}}}
\def\gsim{\raise0.3ex\hbox{$\;>$\kern-0.75em\raise-1.1ex\hbox{$\sim\;$}}}

\begin{document}
\setlength{\unitlength}{1cm}
\setlength{\mathindent}{0cm}
\thispagestyle{empty}
\null
\hfill WUE-ITP-03-016\\
\null
\hfill UWThPh-2003-24\\
\null
\hfill HEPHY-PUB 779/03\\
\null
\hfill hep-ph/0310011\\
\begin{center}
	{\Large \bf Impact of beam polarization on CP asymmetries in 
		neutralino pair production}
\vskip 2.5em
{\large
{\sc A.~Bartl$^{a}$\footnote{e-mail:
        bartl@ap.univie.ac.at}
     H.~Fraas$^{b}$\footnote{e-mail:
		  fraas@physik.uni-wuerzburg.de},
	  T.~Kernreiter$^{a}$\footnote{e-mail:
        tkern@qhepu3.oeaw.ac.at}
	  O.~Kittel$^{a,b}$\footnote{e-mail:
		  kittel@physik.uni-wuerzburg.de},
	  W.~Majerotto$^{c}$\footnote{e-mail:
        majer@qhepu3.oeaw.ac.at}
}}\\
[1ex]
{\normalsize \it
$^{a}$ Institut f\"ur Theoretische Physik, Universit\"at Wien, 
Boltzmanngasse 5, A-1090 Wien, Austria}\\
{\normalsize \it
$^{b}$ Institut f\"ur Theoretische Physik, Universit\"at
W\"urzburg, Am Hubland, D-97074~W\"urzburg, Germany}\\
{\normalsize \it
$^{c}$ Institut f\"ur Hochenergiephysik, \"Osterreichische
Akademie der Wissenschaften, Nikolsdorfergasse 18, 
A-1050 Wien, Austria}\\
\vskip 1em
\end{center} \par

\begin{abstract}

We analyze the dependence of CP-odd asymmetries
on longitudinally polarized $e^+$ and $e^-$ beams   
in neutralino production and subsequent 
leptonic two-body decays. We give numerical examples of the 
asymmetries and of the cross sections 
in the Minimal Supersymmetric  Standard Model with complex parameters 
$\mu$, $M_1$ and $A_{\tau}$ for  a 
collider with $\sqrt{s}=500$ GeV.
We show that longitudinally polarized electron and positron beams can enhance 
considerably both the asymmetries and the cross sections.

\end{abstract}


\section{Introduction}

In the neutralino sector of the Minimal Supersymmetric Standard Model 
(MSSM) \cite{haberkane}, the gaugino mass parameter $M_1$, the higgsino 
mass parameter $\mu$, and the trilinear coupling parameter $A_{\tau}$ 
in the stau sector, can be complex.
The physical phases $\varphi_{M_1}$,  $\varphi_{\mu}$ and 
$\varphi_{A_{\tau}}$ of these parameters can  cause large CP-violating effects 
already at tree level. 

In neutralino production
\begin{eqnarray} \label{production}
   e^-+e^+&\to&
	\tilde{\chi}^0_i+\tilde{\chi}^0_j 
\end{eqnarray}
and the subsequent leptonic two-body decay of one of the neutralinos
\begin{eqnarray} \label{decay_1}
   \tilde{\chi}^0_i&\to& \tilde{\ell} + \ell_1,  
\end{eqnarray}
and of the decay slepton
	\begin{eqnarray} \label{decay_2}
		\tilde{\ell}&\to&\tilde{\chi}^0_1+ \ell_2;\;\;\; \ell_{1,2}= e,\mu,\tau,
\end{eqnarray}
the neutralino spin correlations lead to several CP-odd asymmetries. 
With the triple product 
$	{\mathcal T} = (\vec p_{e^-} \times 
		\vec p_{\ell_2})\cdot \vec p_{\ell_1}$, 
we define the \emph{T-odd} asymmetry of the cross section $\sigma$
for the processes (\ref{production})-(\ref{decay_2}):
\begin{eqnarray} \label{Tasymmetry}
{\mathcal A}_{\rm T} = \frac{\sigma({\mathcal T}>0)
				 -\sigma({\mathcal T}<0)}
					{\sigma({\mathcal T}>0)+
					\sigma({\mathcal T}<0)}.
\end{eqnarray}
If absorbtive phases are neglected, ${\mathcal A}_{\rm T}$ is CP-odd
due to CPT invariance.
The dependence of  ${\mathcal A}_{\rm T}$ on 
$\varphi_{M_1}$ and  $\varphi_{\mu}$ was analyzed 
in \cite{oshimo,choi1,olaf}.

In case the neutralino decays into a $\tau$-lepton,
$\tilde{\chi}^0_i\to\tilde{\tau}_k^{\pm} \tau^{\mp}$,
$k=1,2$,
the T-odd transverse $\tau^-$ and $\tau^+$
polarizations $P_2$ and $\bar P_2$, respectively, 
give rise to the \emph{CP-odd} observable 
\begin{eqnarray} \label{ACP}
{\mathcal A}_{\rm CP}=\frac{1}{2}(P_2-\bar{P}_2),
\end{eqnarray}
which is also sensitive to $\varphi_{A_{\tau}}$.
For various MSSM scenarios, ${\mathcal A}_{\rm CP}$ was discussed 
in \cite{staupol}.
For measuring the asymmetries,
it is crucial to have both large asymmetries and large cross
sections.
In this note we study the impact of  longitudinally 
polarized $e^+$ and $e^-$ beams of a future linear 
collider in the 500 GeV range 
on the asymmetries  ${\mathcal A}_{\rm T}$, ${\mathcal A}_{\rm CP}$
and on the cross sections $\sigma$.

\section{Numerical results
	\label{Numerical results}}

We present numerical results for 
$e^+e^-\to\tilde\chi^0_1 \tilde\chi^0_2$
with the subsequent leptonic decay of $ \tilde\chi^0_2$
for a linear collider with $\sqrt{s}=500$ GeV.
For ${\mathcal A}_{\rm T}$, Eq.~(\ref{Tasymmetry}), 
we study the neutralino 
decay  into the right selectron and right smuon, 
$\tilde \chi^0_2\to\tilde\ell_R\ell_1$, $\ell=e,\mu$ and
for ${\mathcal A}_{\rm CP}$, Eq.~(\ref{ACP}),
that  into the lightest scalar tau,
$\tilde \chi^0_2\to\tilde\tau_1\tau$.
We study the dependence of the asymmetries 
and the cross sections on the beam polarizations
$P_{e^-}$ and $P_{e^+}$ for fixed parameters
$\mu = |\mu| \, e^{ i\,\varphi_{\mu}}$, 
$M_1 = |M_1| \, e^{ i\,\varphi_{M_1}}$, 
$A_{\tau} = |A_{\tau}| \, e^{ i\,\varphi_{A_{\tau}}}$,
$M_2$ and $\tan \beta$. 
We assume $|M_1|=5/3 M_2\tan^2\theta_W $ and use the
renormalization group equations \cite{hall} for the 
selectron and smuon masses,
$m_{\tilde\ell_R}^2 = m_0^2 +0.23 M_2^2
-m_Z^2\cos 2 \beta \sin^2 \theta_W$ with $m_0=100$ GeV.
The interaction Lagrangians and details on stau mixing can
be found in \cite{olaf}.

\begin{figure}[t]
\begin{picture}(10,8)(0,0)
   \put(0,0){\includegraphics{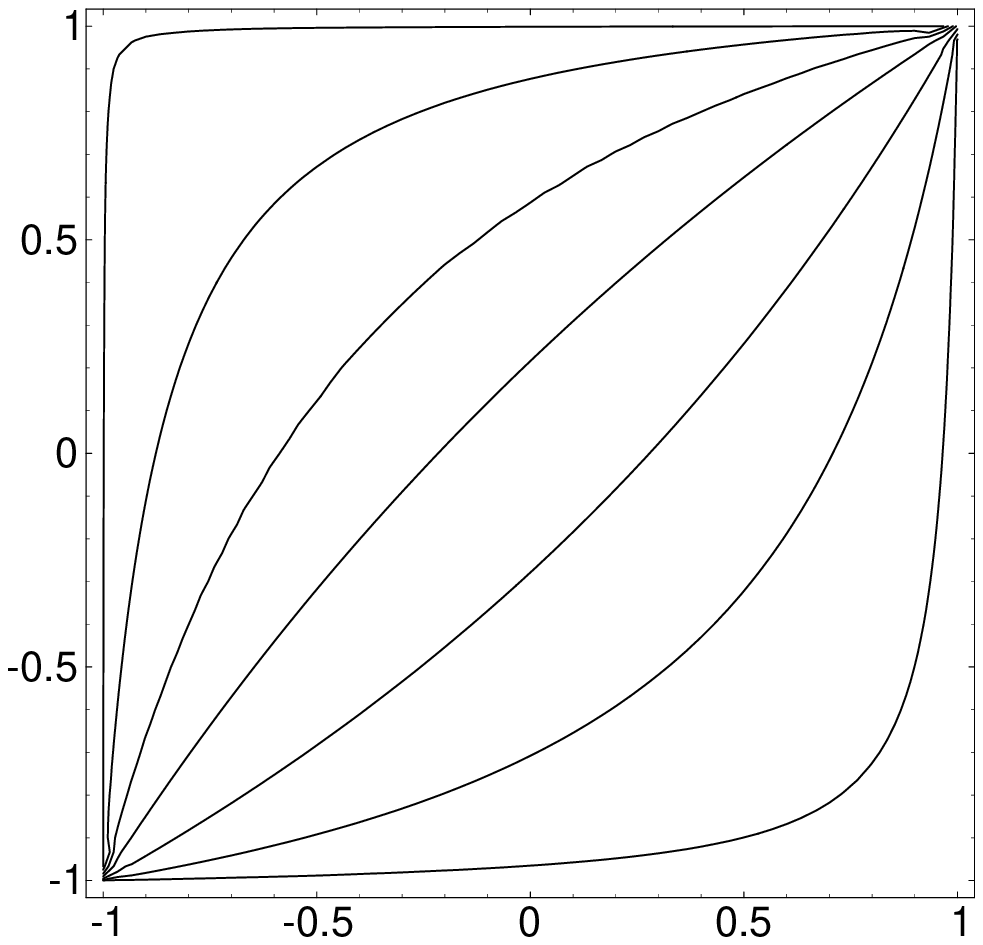}}
	\put(3.,7.5){\fbox{${\mathcal A}_{\rm T}$ in \% }}
	\put(6.5,-.3){$P_{e^-}$}
	\put(0.2,7.3){$P_{e^+}$}
	\put(1.2,6.3){-4.5}
	\put(2.2,5.2){-3}
	\put(3.0,4.4){0}
	\put(3.6,3.8){3}
	\put(4.5,3){6}
	\put(5.2,2.1){8}
	\put(6.5,1.){9}
   \put(0.5,-.3){Fig.~\ref{plot_1}a}
	\put(8,0){\includegraphics{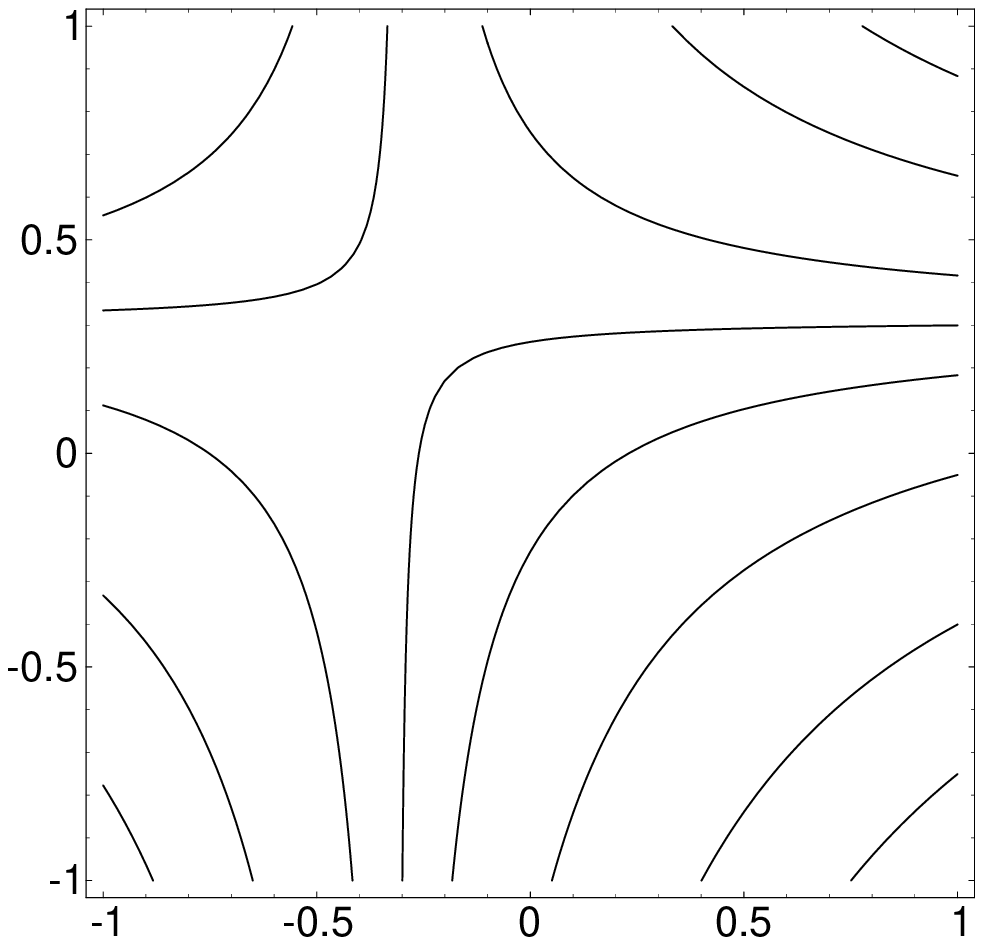}}
	\put(9.5,7.5){\fbox{$\sigma(e^+\,e^- \to\tilde{\chi}^0_1 
				\tilde{\chi}^0_1 \ell_1 \ell_2 )$ in fb}}
	\put(14.5,-.3){$P_{e^-}$}
	\put(8.2,7.3){$P_{e^+}$}
	\put(8.5,-.3){Fig.~\ref{plot_1}b}
	\put(9.2,1.2){4}
	\put(9.5,2.1){12}
	\put(10.0,3.6){20}
	\put(10.2,5.2){24}
	\put(11.5,3.9){24}
	\put(9.3,6.1){28}
	\put(12.3,3.1){28}
	\put(13.2,2.3){36}
	\put(14.,1.5){48}
	\put(14.6,0.8){60}
	\put(11.8,5.3){20}
	\put(13.2,5.8){12}
	\put(14.3,6.3){4}
\end{picture}
\vspace*{.5cm}
\caption{
Contour lines of ${\mathcal A}_{\rm T}$ and $\sigma$ for
$\varphi_{M_1}=0.2\pi $, $\varphi_{\mu}=0$,
$|\mu|=240$ GeV, $M_2=400$ GeV, $\tan \beta=10$ and  $m_0=100$ GeV.
\label{plot_1}}
\end{figure}

In Fig.~\ref{plot_1}a we show the dependence 
of ${\mathcal A}_{\rm T}$ on the beam polarization  
for $\varphi_{M_1}=0.2~\pi $ 
and $\varphi_{A_{\tau}}=\varphi_{\mu}=0$. 
A small value of $\varphi_{\mu}$ is suggested by constraints on
electron and neutron electric dipole moments (EDMs) \cite{edmsexp} 
for a typical SUSY scale of the order of a few 100 GeV
(for a review see, e.g., \cite{edmstheo}).
It is remarkable that in our scenario the asymmetry can be close to 10\% 
even for the small value of $\varphi_{M_1}=0.2~\pi $ and for $\varphi_{\mu}=0$.
The cross section $\sigma=
\sigma(e^+e^-\to\tilde\chi^0_1\tilde\chi^0_2 ) \times
{\rm BR}(\tilde \chi^0_2\to\tilde\ell_R\ell_1)\times
{\rm BR}(\tilde\ell_R\to\tilde\chi^0_1\ell_2)$ is
shown in Fig.~\ref{plot_1}b. For our scenario
with $|A_{\tau}|=250$ GeV and  $\varphi_{A_{\tau}}=0$,
the neutralino branching ratio is
BR$(\tilde \chi^0_2\to\tilde\ell_R\ell_1)$ = 0.63
(summed over both signs of charge) 
and BR($ \tilde\ell_R \to \tilde\chi^0_1\ell_2$) = 1.
Note that  the asymmetry ${\mathcal A}_{\rm T}$ and the cross 
section $\sigma$ are both considerably enhanced for 
negative positron and positive electron beam polarization. 
This choice of polarization enhances the contributions
of the right slepton exchange in the neutralino production,
Eq.~(\ref{production}), and reduces that of left 
slepton exchange \cite{gudi1,gudi2}. 
While the contributions of right and left slepton exchange 
enter $\sigma$ with the 
same sign, they enter
${\mathcal A}_{\rm T}$ with  opposite sign, which accounts for
the sign change of ${\mathcal A}_{\rm T}$.

%
\begin{figure}[t]
\begin{picture}(10,8)(0,0)
   \put(0,0){\includegraphics{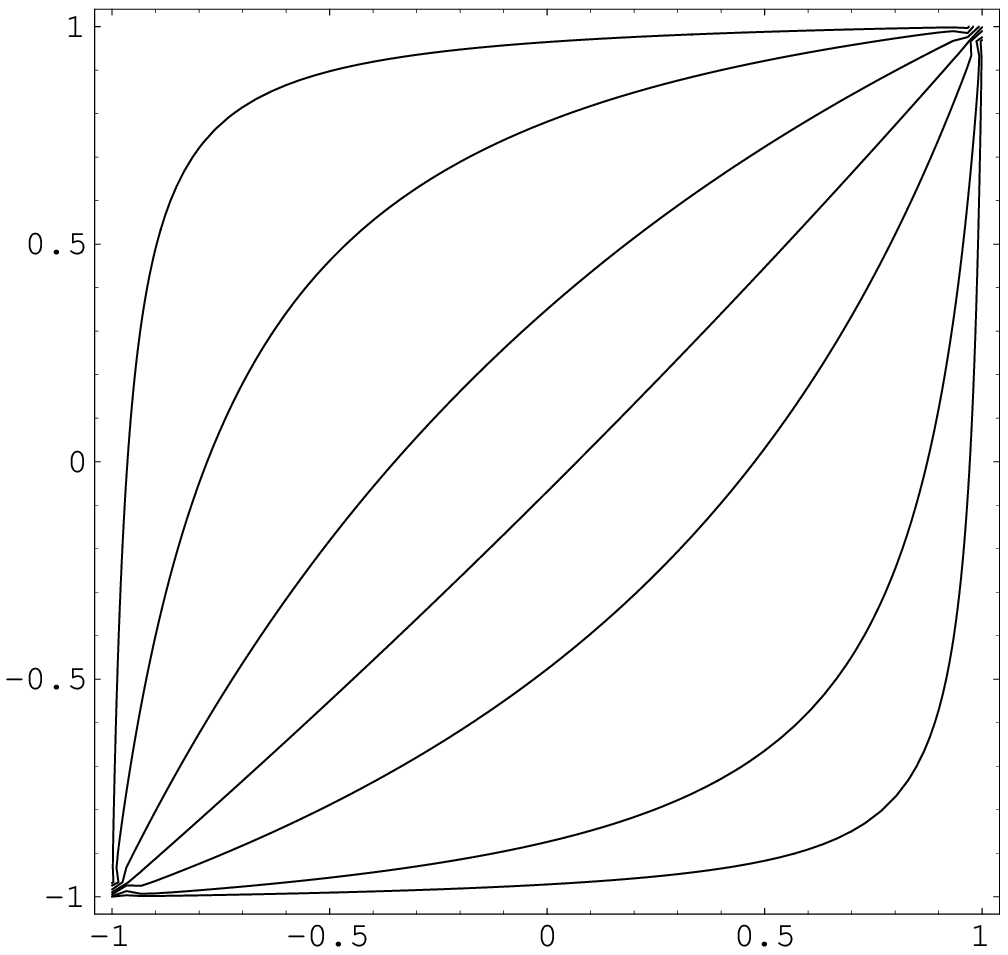}}
	\put(3.,7.5){\fbox{${\mathcal A}_{\rm CP}$ in \% }}
	\put(6.5,-.3){$P_{e^-}$}
	\put(0.2,7.3){$P_{e^+}$}
	\put(1.,6.4){-14.5}
	\put(2.,5.4){-12}
	\put(3.0,4.4){-6}
	\put(3.8,3.7){0}
	\put(4.5,3){6}
	\put(5.3,2.1){12}
	\put(6.3,.8){13.5}
   \put(0.5,-0.3){Fig.~\ref{plot_2}a}
	\put(8,0){\includegraphics{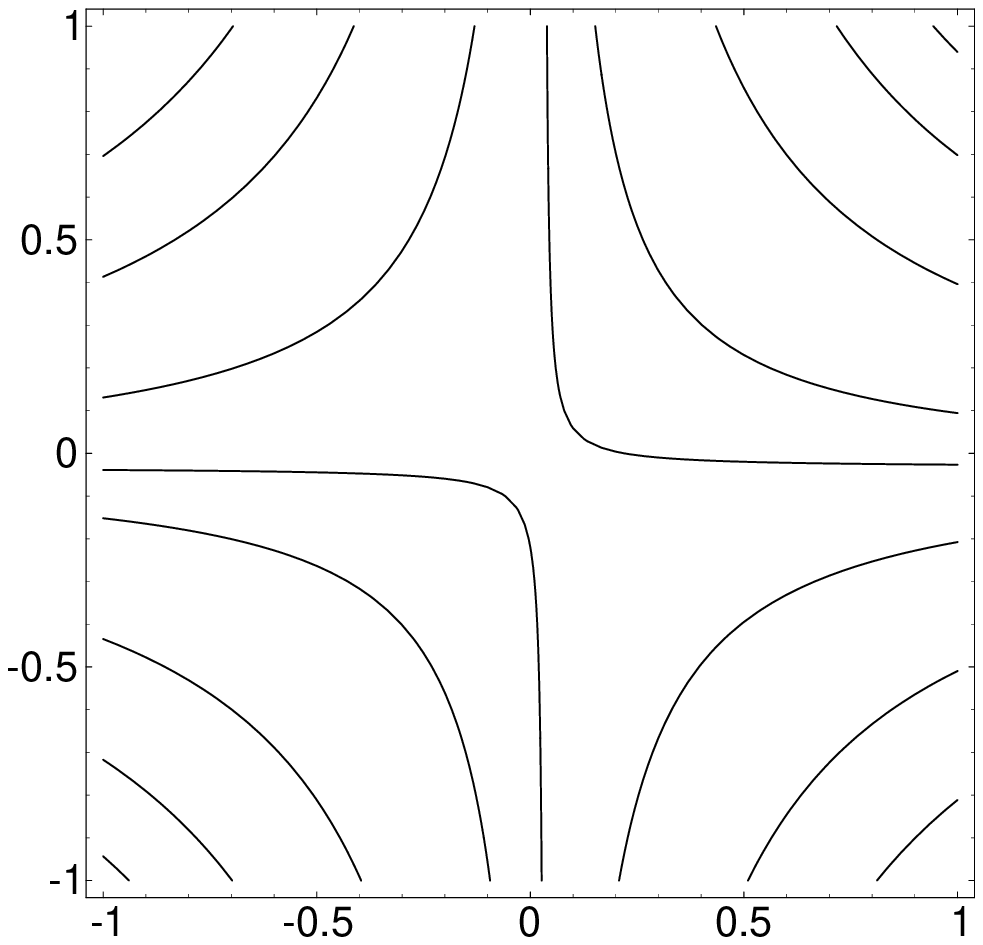}}
	\put(9.5,7.5){\fbox{$\sigma(e^+\,e^- \to\tilde{\chi}^0_1 
				\tilde\tau^+_1 \tau^-)$ in fb}}
	\put(14.5,-.3){$P_{e^-}$}
	\put(8.2,7.3){$P_{e^+}$}
	\put(8.5,-.3){Fig.~\ref{plot_2}b}
	\put(9.1,0.8){1}
	\put(9.6,1.2){5}
	\put(10.2,1.8){10}
	\put(11.0,2.6){15}
	\put(11.1,3.7){17}
	\put(11.,4.7){20}
	\put(10.2,5.5){25}
	\put(9.5,6.0){30}
		\put(13.,3.3){17}
		\put(12.8,4.3){15}
		\put(13.6,5.2){10}
		\put(14.3,5.9){5}
		\put(14.7,6.4){1}
		\put(13.3,2.1){20}
	\put(14.1,1.3){25}
	\put(14.6,0.7){30}
\end{picture}
\vspace*{.5cm}
\caption{
Contour lines   of ${\mathcal A}_{\rm CP}$ and $\sigma$
for $\varphi_{A_{\tau}} =0.5\pi$, $|A_{\tau}|=1500$ GeV,
$\varphi_{M_1}=\varphi_{\mu}=0$,
$|\mu|=250$ GeV, $M_2=200$ GeV, $\tan \beta=5$ and  $m_0=100$ GeV.
\label{plot_2}}
\end{figure}

In Fig.~\ref{plot_2}a we show the contour lines of  the $\tau$
polarization asymmetry ${\mathcal A}_{\rm CP}$, Eq.~(\ref{ACP}), for 
$\varphi_{A_{\tau}} =0.5\pi$ and
$\varphi_{M_1}=\varphi_{\mu}=0$ in the $P_{e^-}$-$P_{e^+}$ plane.
We have chosen a large value of 
$|A_{\tau}|=1500$ GeV because ${\mathcal A}_{\rm CP}$ increases with  
increasing $|A_{\tau}| \gg |\mu|\tan\beta$ \cite{staupol}.
For unpolarized beams the asymmetry is 1\%. However, it 
reaches values of  more than $\pm 13\%$ 
if the  $e^+$ and $e^-$ beams are polarized with the opposite  
sign. 
If at least one of the beams is polarized 
(e.g. $P_{e^-}=0.8$, $P_{e^+}=0.6$), the asymmetries are somewhat
smaller ($\sim 10\%$). 
The reason for this dependence is again the enhancement of 
either the right or the left selectron exchange contributions in the 
production process.
The cross section $\sigma=
\sigma(e^+e^-\to\tilde\chi^0_1\tilde\chi^0_2 ) \times
{\rm BR}(\tilde \chi^0_2\to\tilde\tau_1^+\tau^-)$ is shown
in Fig.~\ref{plot_2}b with 
${\rm BR}(\tilde \chi^0_2\to\tilde\tau_1^+\tau^-)=0.22$.
Also  $\sigma$ is very sensitive to variations of the 
beam polarization and varies between 1 fb and 30 fb.

\begin{figure}[t]
\begin{picture}(10,8)(0,0)
   \put(0,0){\includegraphics{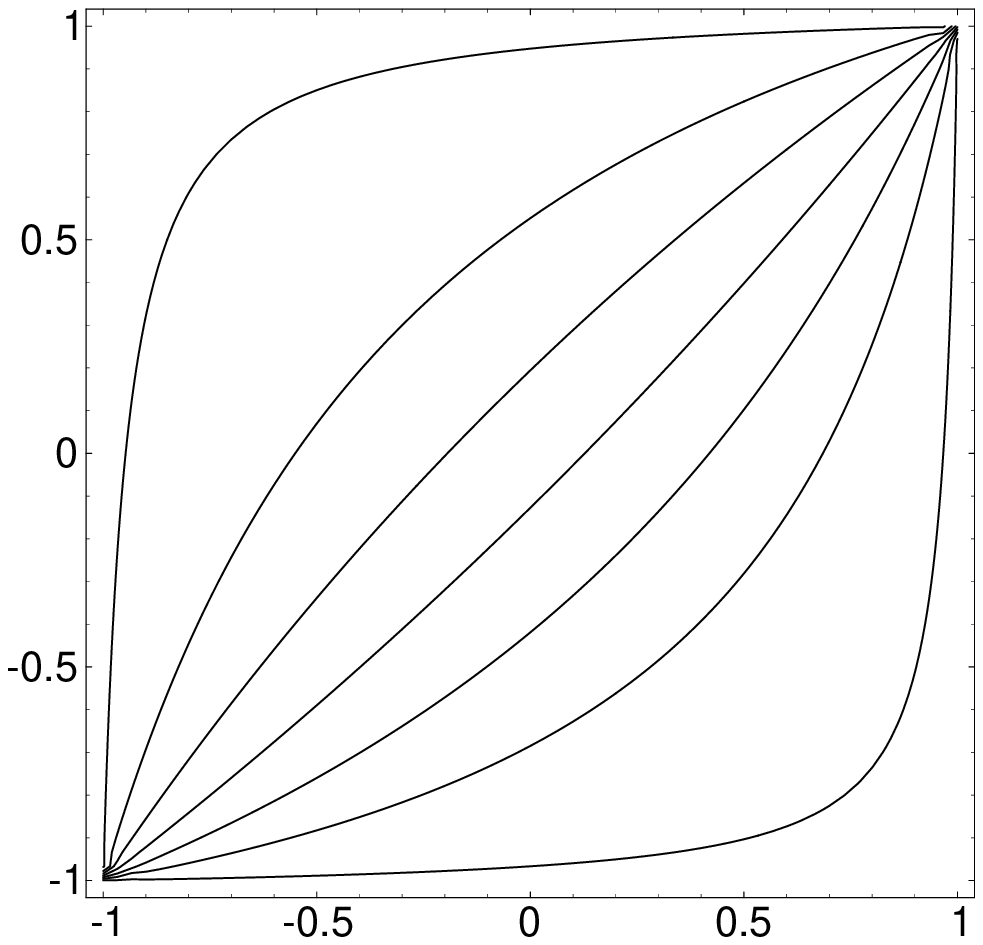}}
	\put(3.,7.5){\fbox{${\mathcal A}_{\rm CP}$ in \% }}
	\put(6.5,-.3){$P_{e^-}$}
	\put(0.2,7.3){$P_{e^+}$}
		\put(1.4,6.1){-12}
	\put(2.9,5.){-9}
	\put(3.4,4.3){-6}
	\put(3.9,3.7){-3}
	\put(4.5,3.2){0}
	\put(5.3,2.2){3}
	\put(6.4,.9){6.5}
   \put(0.5,-.3){Fig.~\ref{plot_3}a}
	\put(8,0){\includegraphics{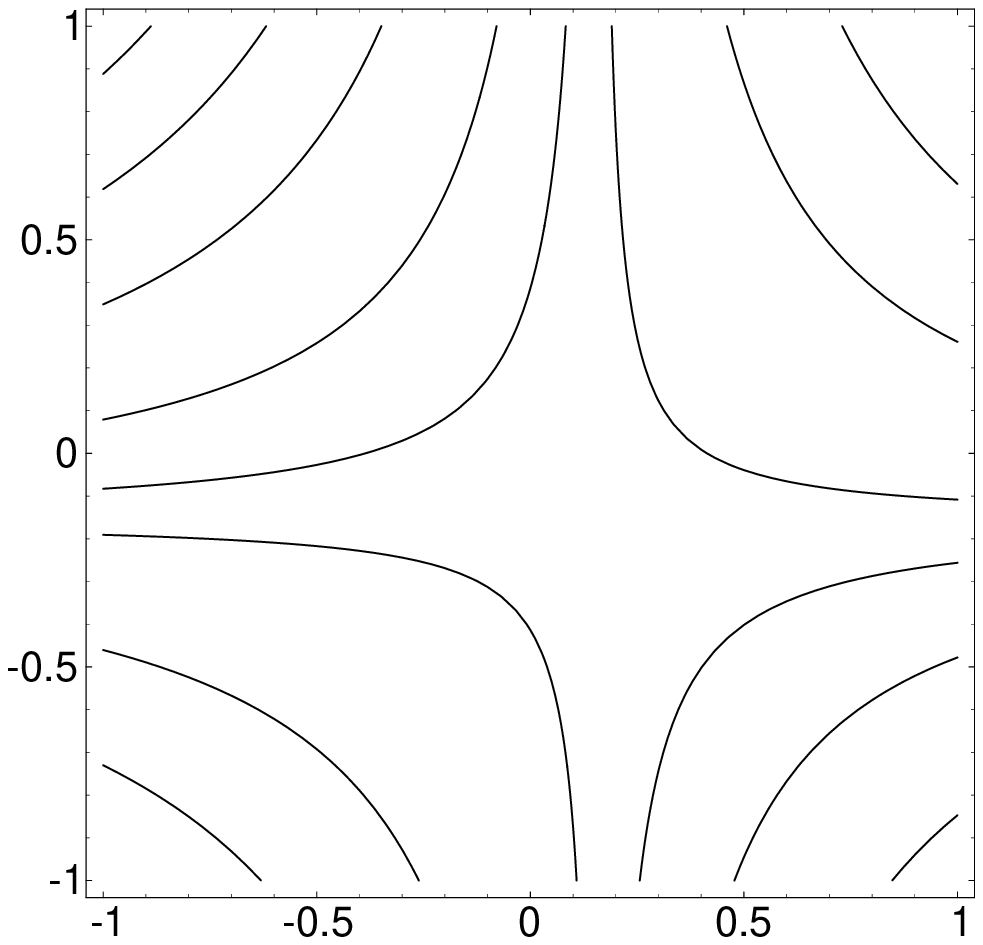}}
	\put(9.5,7.5){\fbox{$\sigma(e^+\,e^- \to\tilde{\chi}^0_1 
				\tilde\tau^+_1 \tau^-)$ in fb}}
	\put(14.5,-.3){$P_{e^-}$}
	\put(8.2,7.3){$P_{e^+}$}
	\put(8.5,-.3){Fig.~\ref{plot_3}b}
	\put(9.6,1.2){5}
	\put(10.4,1.7){10}
	\put(11.4,2.1){15}
	\put(11.6,3.8){17}
	\put(10.9,4.6){20}
	\put(10.1,5.2){25}
	\put(9.5,5.7){30}
	\put(9.1,6.3){35}
		\put(12.9,4.3){15}
		\put(13.6,5.){10}
		\put(14.3,5.9){5}
		\put(13.4,2.1){17}
	\put(14.1,1.4){20}
	\put(14.7,0.7){25}
\end{picture}
\vspace*{.5cm}
\caption{
Contour lines of ${\mathcal A}_{\rm CP}$ and $\sigma$ for  
$\varphi_{M_1}= 0.2\pi$, $\varphi_{\mu}=0$,
$|\mu|=250$ GeV, $M_2=200$ GeV, 
$\varphi_{A_{\tau}} =0$, $|A_{\tau}|=250$ GeV,
$\tan \beta=5$ and  $m_0=100$ GeV.
\label{plot_3}}
\end{figure}

Since the asymmetry ${\mathcal A}_{\rm CP}$ is also very sensitive to
the phases $\varphi_{M_1}$ and $\varphi_{\mu}$ we show
for $\varphi_{M_1}=0.2\pi$ and 
$\varphi_{\mu}=\varphi_{A_{\tau}}=0$,
the dependence of ${\mathcal A}_{\rm CP}$ and $\sigma= 
\sigma(e^+e^-\to\tilde\chi^0_1\tilde\chi^0_2 ) \times
{\rm BR}(\tilde \chi^0_2\to\tilde\tau_1^+\tau^-)$
on the beam polarization in Figs.~\ref{plot_3}a, b, respectively.
The neutralino branching ratio is 
${\rm BR}(\tilde\chi^0_2\to\tilde\tau_1^+\tau^-)=0.19$
for our scenario.
Despite the small phases, 
${\mathcal A}_{\rm CP}$ reaches values up to $-12\%$  
for negative $e^-$ and positive $e^+$ beam polarizations.

\section{Summary and conclusion
	\label{Summary and conclusion}}

Within the MSSM we have analyzed the dependence on the beam polarization
of two CP-odd asymmetries in 
$e^+e^- \to\tilde\chi^0_1  \tilde\chi^0_2$
and the subsequent leptonic two-body decay of $\tilde\chi^0_2$.
For the decay process 
$\tilde\chi^0_2 \to \tilde\ell_R \ell_1$,
$\tilde\ell_R \to \tilde\chi^0_1 \ell_2$ with $ \ell_{1,2}= e,\mu$,
we have found that the asymmetry ${\mathcal A}_{{\rm T}}$ of 
the triple product
$ (\vec p_{e^-}\times\vec p_{\ell_2})\cdot \vec p_{\ell_1}$,
which is sensitive to $\varphi_{M_1}$ and $\varphi_{\mu}$,
can be twice as large if polarized beams are used, with e.g.
$P_{e^-}=0.8$ and $P_{e^+}=-0.6$.  Also for these polarizations the
cross section can be enhanced up to a factor of 2. 
For the neutralino decay, 
$\tilde\chi^0_2 \to \tilde\tau_1^{\mp}\tau^{\pm}$,
we have given numerical examples for the beam polarization dependence of the
CP-odd $\tau$ polarization asymmetry ${\mathcal A}_{\rm CP}$,
which is also sensitive to $\varphi_{A_{\tau}}$.
For the scenarios considered, 
both ${\mathcal A}_{\rm CP}$ and the cross section  depend sensitively 
on the beam polarizations and can be enhanced by a factor 
between 2 and 3. The dependence on the beam polarizations of the asymmetries 
${\mathcal A}_{\rm T}$, ${\mathcal A}_{\rm CP}$ and of 
the cross sections
is due to the contributions from right and left selectron exchange 
in the neutralino production process. Generally, negative (positive) 
$e^-$ and positive (negative ) $e^+$ beam polarization enhances 
right (left) selectron exchange.
Due to the fact that both asymmetries and cross 
sections can be enhanced significantly, we conclude
that the option of having both beams polarized at an $e^+e^-$-collider
is advantageous for the determination of the CP-odd asymmetries.

\section{Acknowledgement}
This work was supported by the `Fonds zur
F\"orderung der wissenschaftlichen Forschung' (FWF) of Austria, projects
No. P13139-PHY and No. P16592-N02 and by the European Community's
Human Potential Programme under contract HPRN-CT-2000-00149.
This work was also supported by the 'Deutsche Forschungsgemeinschaft'
(DFG) under contract Fr 1064/5-1.

\end{document}